# *Ab initio* study of the elastic and electronic properties of tetragonal Th$_2$NiC$_2$

**I.R. Shein, A.L. Ivanovskii**

*Institute of Solid State Chemistry, Ural Branch of the Russian Academy of Sciences, 620990 Ekaterinburg, Russia*
*E-mail address:* shein@ihim.uran.ru

This work reports on the elastic and electronic properties of the newly discovered superconductor Th$_2$NiC$_2$ (*A .Machado, et al., Supercond. Sci. Technol.* 25 (2012) 045010) as obtained within *ab initio* calculations. We found that Th$_2$NiC$_2$ is mechanically stable and it will behave as a ductile material exhibiting enhanced elastic anisotropy in shear and a rather low hardness Our data reveal that for Th$_2$NiC$_2$ the Fermi level is located in a deep DOS minimum and the experimentally observed increase in $T_C$ in the sequence Th$_2$NiC$_2$ → Th$_{1.8}$Sc$_{0.2}$NiC$_2$ may be explained by the growth of $N(E_F)$. We also speculate that (i) an increase in the hole concentration will promote exchange splitting of Ni $3d^{\uparrow\downarrow}$ bands, therefore the hole-doped Th$_2$NiC$_2$ should have a certain concentration border, where a phase transition from the superconducting to the magnetic state will be expected, and (ii) an increase in $N(E_F)$ (and, probably, in $T_C$) for Th$_2$NiC$_2$-based materials may be also achieved by an alternative way: by electron doping - for example, by partial substitution of V for Th or Cu for Ni, as well as by partial substitution of N for C with the formation of Th-Ni carbonitrides like Th$_2$NiC$_{2-x}$N$_x$.

**1. Introduction**

The coexistence (or competition) of superconductivity and magnetic spin-fluctuations is one of the exciting issues of solid state. Here, Ni-based superconductors (SCs) have attracted a great interest, since many of such materials lie in close proximity to magnetism - for example, MgCNi$_3$ and some other Ni-rich ternary antiperovskite-like carbides and nitrides [1-12].

Recently, research was concentrated on more complex systems, which involve simultaneously Ni and *d* or *f* metals. Among them there are so-called Heusler superconductors (like ZrNi$_2$Ga, $T_C$ ~ 2.9 K [13]) and a broad family of layered ternary and quaternary materials such as GdNiBiO ($T_C$ ~ 4.5 K), La$_3$Ni$_4$Si$_4$ ($T_C$ ~ 1.0 K), and La$_3$Ni$_4$P$_4$O$_2$ ($T_C$ ~ 2.2 K), which comprise building blocks Ni$_2$X$_2$ (X = As, P, Bi, Si, Ge, B) as common structural elements, see review [14]. Besides, new five-component Ni-based pnictide oxides (Ni$_2$Pn$_2$)(Sr$_4$M$_2$O$_6$) (Pn = P, As; M = Sc, V) with $T_C$ ~ 2.6 – 3.7 K have been discovered and examined [15-17].

Very recently, this family of Ni-containing SCs was expanded by two interesting ternary carbides: Th$_3$Ni$_5$C$_5$ and Th$_2$NiC$_2$, for which a transition to the superconducting state was found at $T_C$ ~ 5.0 K and $T_C$ ~ 8.5 K, respectively [18,19]. Moreover, it was found that partial substitution of Sc for Th in Th$_2$NiC$_2$ (*i.e.* hole doping) induces an increase in $T_C$ to ~ 11.2 K for the composition Th$_{1.8}$Sc$_{0.2}$NiC$_2$ [19].

In view of these circumstances, in this work we have performed first-principle calculations for Th$_2$NiC$_2$. Our main goal was to study the electronic band structure and inter-atomic bonding picture for this new superconducting phase, as well as to evaluate the structural and elastic parameters (elastic constants, bulk, shear, Young's moduli, Poisson's ratio *etc.*) of $Th_2NiC_2$, which are analyzed in comparison with cubic ThC.

## 2. Models and computational aspects

The considered phase $Th_2NiC_2$ possesses [19,20] a tetragonal structure (space group *I*4/*mmm*, # 139), where the [$Th_2C_2$] blocks are sandwiched between Ni atomic sheets, Fig. 1. The Wyckoff positions of atoms are listed in Table 1.

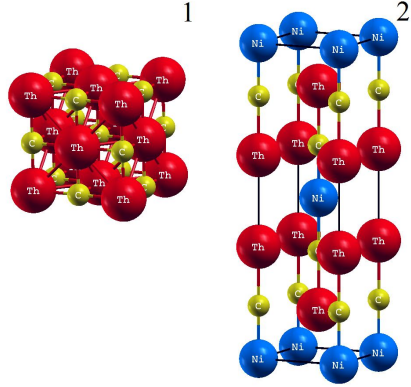

**Fig. 1.** Crystal structures of cubic ThC (1) and tetragonal $Th_2NiC_2$ (2)

**Table 1** Optimized atomic coordinates for $Th_2NiC_2$ in comparison with experiment

| atom | x | y | z * |
|---|---|---|---|
| Th | 0 | 0 | 0.37727 (0.35402) |
| Ni | 0 | 0 | 0 |
| C | 0 | 0 | 0.15607 (0.156) |

* available experimental data, Ref. [20], are given in parentheses.

Our calculations were performed by means of the full-potential method with mixed basis APW+lo (LAPW) implemented in the WIEN2k suite of programs [21]. Exchange and correlation were also described in the form of GGA [22]. The basis set inside each MT sphere was split into core and valence subsets. The core states were treated within the spherical part of the potential only, and were assumed to have a spherically symmetric charge density confined within MT spheres. The valence part was treated with the potential expanded into spherical harmonics to $l$ = 4. The valence wave functions inside the spheres were expanded to $l$ = 10. The plane-wave expansion with $R_{MT} \times K_{MAX}$ was equal to 7.0, and $k$ sampling with 12×12×12 *k*-points mesh in the full Brillouin zone was used. The MT sphere radii were chosen to be 2.30 a.u. for Th, 1.85 a.u. for Ni, and 1.70 a.u. for carbon. Relativistic effects were taken into account within the scalar-relativistic treatment including spin-orbit coupling (SOC) as calculated using a second-variational treatment [23]. The self-consistent calculations were considered to converge when the difference in the total energy of the crystal did not exceed 0.1 mRy and the difference in the atomic forces did not exceed 1 mRy/a.u. as calculated at consecutive steps.

Since the ternary phase $Th_2NiC_2$ may be described as a stacking of NaCl-type [$Th_2C_2$] blocks and sheets of Ni atoms, Fig. 1, it seems useful to discuss the examined properties of $Th_2NiC_2$ in comparison with the same for cubic ThC. For this goal we will use the data for cubic thorium monocarbide (c-ThC) obtained by the authors [24] at the same computational level.

## 3. Results and discussion

### 3.1. Structural properties

As the first step, the equilibrium structural parameters: atomic positions and lattice constants ($a$ and $c$) were calculated. The obtained results are presented in Tables 1 and 2, and are in reasonable agreement with the available experiments [19,20].

**Table 2.** Calculated lattice constants ($a$ and $c$, in Å), $c/a$ ratio, and selected inter-atomic distances ($d$, in Å) for $Th_2NiC_2$ in comparison with experiments and structural data for cubic ThC

| phase | results | $a$ | $c$ | $c/a$ | $d$ ** |
|---|---|---|---|---|---|
| $Th_2NiC_2$ | our data | 3.7727 | 12.327 | 3.267 | $d^{Th-Th}$= 3.6041; $d^{Th-C}$= 2.6705; $d^{C-C}$= 3.5326; $d^{Ni-Th}$= 3.2193; $d^{Ni-C}$= 1.9239 |
| | exp. Ref. [19] | 3.748 | 12.349 | 3.294 | |
| | exp. Ref. [20] | 3.750 | 12.340 | 3.291 | |
| ThC | our data, Ref. [24] | 5.3878 | - | - | $d^{Th-Th}$= 3.8097 |
| | theor. Ref. [25] | 5.335-5.344 * | - | - | $d^{Th-C}$= 2.6939 |
| | exp. Ref. [26] | 5.261-5.363 | - | - | $d^{C-C}$= 3.8097 |

* as calculated for various exchange potentials, Ref. [25].
** our data.

The evaluated inter-atomic distances (Table 2) allow us to draw some preliminary conclusions about the chemical bonding picture. For c-ThC, the inter-atomic distances $d$(Th–C) (Table 2) are close to a sum of covalent radii of Th ($R^{Th}$ = 1.65 Å) and carbon ($R^C$ = 0.77 Å), and in this monocarbide an isotropic system of strong covalent Th–C bonds is formed [27,28]. For $Th_2NiC_2$, the values of $d$(Th–C) are by 0.02 Å smaller than for c-ThC, i.e. similar directional covalent Th–C bonds should be expected. The values of $d$(Ni–C) = 1.9239 Å are also close to the sum of $R^{Ni}$ = 1.21 Å and $R^C$ = 0.77 Å, whereas all other types of inter-atomic distances become much greater. Therefore for $Th_2NiC_2$ two types of strong directional bonds should be supposed: covalent Th-C bonds (inside [$Th_2C_2$] blocks) together with covalent Ni-C bonds (along the $z$ axis). Further we will check up this simplified picture using the results of our calculations.

### 3.2. Elastic properties

Let us discuss the elastic properties of $Th_2NiC_2$. The standard "volume-conserving" technique was used in the calculations of stress tensors on strains applied to the equilibrium structure to obtain the elastic constants $C_{ij}$, see review [29], and the values of six independent elastic constants for tetragonal crystal ($C_{11}$, $C_{12}$, $C_{13}$, $C_{33}$, $C_{44}$ = $C_{22}$ and $C_{66}$) were estimated, Table 3. For comparison, in Table 3 the elastic parameters for c-ThC are given, which are evaluated using the values of $C_{ij}$ from our work [24].

**Table 3.** Calculated elastic constants ($C_{ij}$, in GPa), bulk modulus ($B$, in GPa), compressibility ($β$, in GPa$^{-1}$), shear modulus ($G$, in GPa), Pugh's indicator ($G/B$), Young's modulus ($Y$, in GPa), Poisson's ratio ($ν$), elastic anisotropy indexes ($A^U$, $A_B$, and $A_G$), and Vickers hardness ($H_V$, in GPa) for Th$_2$NiC$_2$ in comparison with cubic ThC

| parameter * | Th$_2$NiC$_2$ | ThC | parameter * | Th$_2$NiC$_2$ | ThC |
|---|---|---|---|---|---|
| $C_{11}$ | 233.6 | 162.9 | $G_V$ | 59.3 | 51.1 |
| $C_{12}$ | 97.5 | 69.6 | $G_R$ | 39.5 | 42.8 |
| $C_{13}$ | 54.1 | - | $G_{VRH}$ | 49.4 | 47.0 |
| $C_{33}$ | 285.7 | - | $G/B$ | 0.41 | 0.47 |
| $C_{44}$ | 23.0 | 54.1 | $Y$ | 130.3 | 122.0 |
| $C_{66}$ | 68.1 | - | $ν$ | 0.3187 | 0.2981 |
| $B_V$ | 129.4 | 100.7 | $A^U$ | 2.67 | 0.97 |
| $B_R$ | 110.2 | 100.7 | $A_B$ | 8.0 % | 0% |
| $B_{VRH}$ | 119.8 | 100.7 | $A_G$ | 20.1% | 9% |
| $β$ | 0.00835 | 0.00993 | $H_V$ | 5.8 | 5.4 |

* the subscripts *V*, *R* and *VRH* indicate the Voigt, Reuss, and Voigt–Reuss–Hill approximations, respectively.

First of all, we see that the constants satisfy the criteria of mechanical stability for tetragonal crystals [30]: $C_{11} > 0$, $C_{33} > 0$, $C_{44} > 0$, $C_{66} > 0$, $(C_{11} - C_{12}) > 0$, $(C_{11} + C_{33} - 2C_{13}) > 0$, and $\{2(C_{11} + C_{12}) + C_{33} + 4C_{13}\} > 0$; thus, we can assert that Th$_2$NiC$_2$ is intrinsically stable.

The calculated constants $C_{ij}$ allowed us to estimate the macroscopic elastic moduli of Th$_2$NiC$_2$: monocrystalline bulk ($B$) and shear moduli ($G$) in Voigt ($V$) and Reuss ($R$) approximations, as well as the same parameters for the corresponding polycrystalline state within the Voigt-Reuss-Hill (*VRH*) approximation, see details in [29]. The obtained values (Table 3) show that $B > G$, *i.e.* the parameter limiting the mechanical stability of this material is the shear modulus $G$. Besides, the $G/B$ ratio was proposed [30] as an empirical malleability measure of polycrystalline materials: if $G/B < 0.5$, a material behaves in a ductile manner, and *vice versa*, if $G/B > 0.5$, a material demonstrates brittleness. In our case, according to this indicator ($G/B \sim 0.4$), Th$_2$NiC$_2$ will behave as a ductile material.

The obtained values show that $B^{Th_2NiC_2} > B^{ThC}$ and $G^{Th_2NiC_2} > G^{ThC}$. As the bulk moduli represent the resistance to volume change and the shear moduli $G$ represent the resistance to shear deformation against external forces, an increased average bond strength can be assumed for Th$_2$NiC$_2$.

The Young's modulus ($Y$) is defined as a ratio of linear stress and linear strain, which tells about material's stiffness. The average Young's moduli (estimated as $Y = 9B/\{1 + (3B/G)\}$, see Table 3) demonstrate that Th$_2$NiC$_2$ adopts higher stiffness than ThC. The possible reasons are: the strengthening of Th-C bonds (inside [Th$_2$C$_2$] blocks owing to some reduction of Th-C distances) and the appearance of strong covalent Ni-C bonds, see also below.

Another interesting point regarding the elastic properties is elastic anisotropy (EA), which is related to different bonding character in different crystallographic directions, and correlates with the possibility to induce microcracks in materials. Today, a set of approaches is proposed to estimate EA numerically. Here we used the following relations:

$$A^U = 5G_V/G_R + B_V/B_R - 6$$
$$A_B = (B_V - B_R)/(B_V + B_R),$$
$$A_G = (G_V - G_R)/(G_V + G_R).$$

The first relation presents so-called universal anisotropy index $A^U$; for isotropic crystals $A^U = 0$ and deviations of $A^U$ from zero define the extent of EA [33]. Two other indexes allow us to estimate elastic anisotropy in compressibility ($A_B$) and shear ($A_G$) [34]. The values of $A^U$, $A_B$, and $A_G$ were obtained (Table 3), and they point to considerable EA for this tetragonal phase. We see that $G_V/G_R$ (1.50) > $B_V/B_R$ (1.17), *i.e.* the difference between $G_V$ and $G_R$ affects $A^U$ much more significantly than the difference between $B_V$ and $B_R$. Accordingly, the anisotropy in shear became much higher than in compressibility. This is not surprising taking into account the layered structure of this material and appreciable anisotropy of inter-atomic bonding, see below. Certainly, for c-ThC these indexes are much smaller, Table 3.

Finally, we used the empirical correlation between Vickers hardness ($H_V$) and polycrystalline shear modulus (in Teter's form [35]:  $H_V = 0.1769G$ - 2.899) to roughly evaluate the value of this parameter for the examined material. The results (Table 3) demonstrate that $Th_2NiC_2$ will exhibit a rather low hardness.

*3.3. Electronic properties and inter-atomic bonding*

The electronic bands of $Th_2NiC_2$ are depicted in Fig. 2 in comparison with c-ThC. We see that as going from c-ThC to tetragonal $Th_2NiC_2$ a new manifold of Ni $3d$-like bands arises in the near-Fermi region (from - 4.6 eV to $E_F$ = 0 eV). Besides, the splitting of C $2p$ bands (hybridized with Th- and Ni-like bands) and quasi-core C $2s$ bands (located from - 9.9 eV to - 6.8 eV) is due to crystal symmetry reduction ($Fm3m \rightarrow I4/mmm$) and the formation of additional Ni-C bonds.

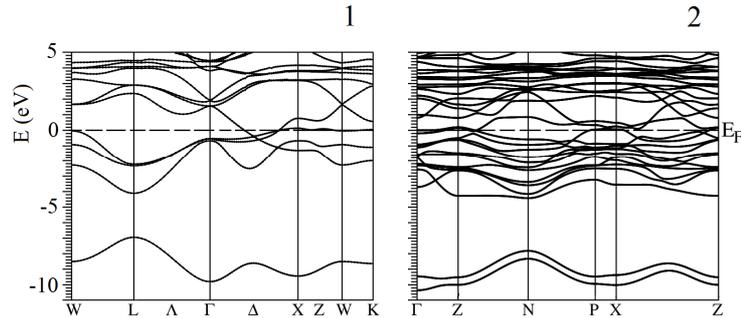

**Fig. 2.** Electronic bands for cubic ThC (1) and tetragonal $Th_2NiC_2$ (2)

The total and atomic-resolved *l*-projected densities of states (DOSs) for c-ThC and $Th_2NiC_2$ phases are depicted in Fig. 3; the total and orbital-decomposed partial DOSs at the Fermi level, $N(E_F)$, are listed in Table 4. From these data we draw the following conclusions.

The main feature of the $Th_2NiC_2$ spectrum (in comparison with c-ThC) is the presence of the band of occupied Ni $3d$ states near the Fermi level, peak B Fig. 3).

These Ni 3*d* states overlap with the C 2*p* states and are responsible for the formation of Ni-C covalent bonds. Besides, from the atomic-resolved *l*-projected DOSs (Fig. 3), where the hybridization of the Th (*d,f*) and C 2*p* states, is well visible, comes the conclusion about Th-C covalent bonding in $Th_2NiC_2$.

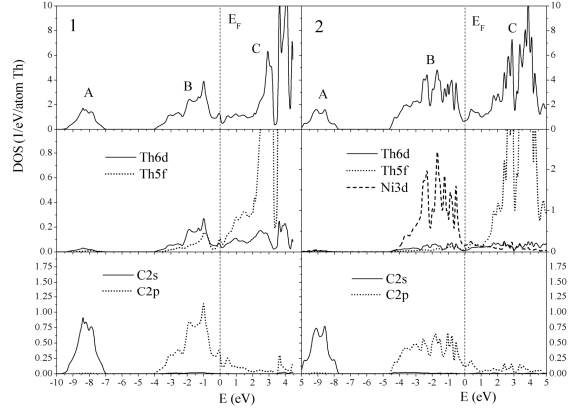

**Fig. 3.** Total and partial densities of state for cubic ThC (1) and tetragonal $Th_2NiC_2$ (2)

The formation of the system of these directional bonds is illustrated using charge density maps, Fig. 4. We see the presence of directional Th-C bonds inside $[Th_2C_2]$ blocks (similar to those for c-ThC, Fig. 4), together with directional Ni-C bonds (in -C-Ni-C- chains located along the *z* axis). Simultaneously we see the absence of directional bonds C-C, Ni-Ni, and Th-Th. As to ionic bonds, the contribution of the Ni 3*d* states to the conduction zone (peak C Fig. 3) is quite small, confirming the qualitative assumption [19] that the nickel configuration is close to $d^{10}$. Therefore, taking the usual oxidation states for $Th^{4+}$ and $C^{4-}$, the zero oxidation state for nickel should be assumed [19]. This means that the bonding picture for the examined phase can be described as an anisotropic mixture of ionic and covalent contributions, where inside neutral $[Th_2C_2]^0$ blocks, the covalent Th-C together with ionic Th-C bonds (owing to charge transfer Th → C) occur; in turn, these blocks are bonded with Ni atomic sheets *via* Ni-C covalent interactions.

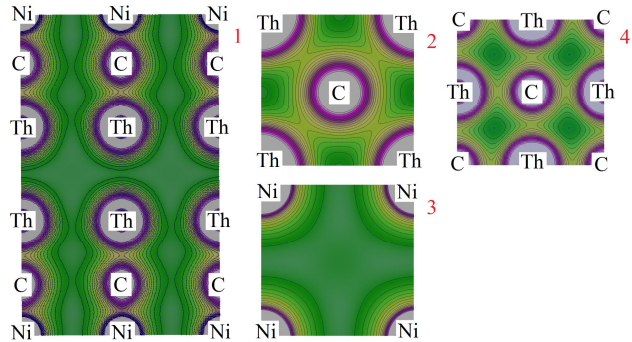

**Fig. 4**. Valence charge density in characteristic planes of $Th_2NiC_2$ (1-3), which illustrate the directional Th-C, and Ni-C bonds in $[Th_2C_2]$ blocks, and between these blocks and Ni sheets (1,2), and the absence of Ni-Ni bonds in Ni sheets (3). The directional Th-C bonds for c-ThC are also shown (4).

The Fermi level for $Th_2NiC_2$ is located in a deep DOS minimum – in agreement with the experiment [19], and the main contributions to $N(E_F)$ arise from the Ni 3$d$ and C 2$p$ states, with some admixtures of Th 5$f$ and Th 6$d$ states. Therefore we speculate that the experimentally probed hole doping [19] should lead to an increase in $N(E_F)$, which becomes a key factor for the increase in $T_C$ in the sequence Th$_2$NiC$_2$ → Th$_{1.8}$Sc$_{0.2}$NiC$_2$. Using a simple rigid-band model, we found that the value of $N(E_F)$ increases from 0.707 states/eV·atom Th (for Th$_2$NiC$_2$) to 2.21 states/eV·atom Th (for Th$_{1.8}$Sc$_{0.2}$NiC$_2$).

On the other hand, the increase in the hole concentration (*i.e.* further emptying of the valence band) will lead to fast growth of the $N^{Ni3d}(E_F)$ value. This will promote exchange splitting of the Ni 3$d^{↑↓}$ bands, leading to magnetic instability of the system. This situation resembles the same for the aforementioned MgCNi$_3$ and some related antiperovskite-like carbides and nitrides [1-12]. Therefore, the hole-doped Th$_2$NiC$_2$ should have a certain concentration border, where a phase transition from the superconducting to the magnetic state should be expected. This issue requires a separate research.

Looking at the DOS profile, Fig. 3, we can suppose that the increase in $N(E_F)$ (and, probably, in $T_C$) may be achieved in the opposite way, namely, by electron doping, *i.e.* by partial filling of the lowest anti-bonding states. This may be realized, for example, by partial substitution of V for Th or Cu for Ni. Another way implies partial substitution of N for C with the formation of Th-Ni carbonirtides like Th$_2$NiC$_{2-x}$N$_x$. Note that in the Th–C–N system a set of ternary carbonitride phases is known; moreover, the ThC$_{1-x}$N$_x$ solid solutions are superconducting with the upper critical transition at $T_C$ ~ 5.8 K for the composition ThC$_{0.78}$N$_{0.22}$ [26,35].

## 4. Conclusions

In the present paper, the structural, elastic, electronic properties and inter-atomic bonding for the newly discovered superconductor Th$_2$NiC$_2$ were examined by means of the first-principle calculations (FLAPW approach, where the relativistic effects were taken into account within the scalar-relativistic treatment including spin-orbit coupling (SOC)) and were discussed in comparison with c-ThC.

We found that Th$_2$NiC$_2$ is mechanically stable, and the parameter limiting the mechanical stability of this material is the shear modulus $G$. Besides, Th$_2$NiC$_2$ will behave as a ductile material, exhibit enhanced elastic anisotropy in shear (greater than in compressibility), and will adopt a rather low hardness.

The picture of chemical bonding in Th$_2$NiC$_2$ can be described as an anisotropic mixture of covalent and ionic contributions, where inside [Th$_2$C$_2$] blocks covalent Th-C bonds emerge, and these blocks are bonded with the Ni atomic sheets through directional Ni-C bonds. Besides, the inter-atomic ionic interactions occur owing to charge transfer Th → C.

Our data reveal that for Th$_2$NiC$_2$ the Fermi level is located in a deep DOS minimum; therefore the experimentally obtained increase in $T_C$ in the sequence Th$_2$NiC$_2$ → Th$_{1.8}$Sc$_{0.2}$NiC$_2$ [19] may be explained by the growth of $N(E_F)$. On the other hand, the increase in the hole concentration will lead to fast growth of $N^{Ni3d}(E_F)$ promoting exchange splitting of the Ni 3$d^{↑↓}$ bands. As a result, the hole-doped Th$_2$NiC$_2$ should have a certain concentration border, where the phase transition from the superconducting to the magnetic state should be expected.

Finally, we speculate that the increase in $N(E_F)$ (and, probably, in $T_C$) for Th$_2$NiC$_2$-based materials may be achieved by electron doping – for example, by partial substitution of V for Th or Cu for Ni, as well as by partial substitution of N for C with the formation of Th-Ni carbonirtides like Th$_2$NiC$_{2-x}$N$_x$. The corresponding experiments seem very desirable.

**References**

[1] T. He, Q. Huang, A.P. Ramirez, Y. Wang, K.A. Regan, N. Rogado, M.A. Hayward, M.K. Haas, J.S. Slusky, K. Inumara, H.W. Zandbergen, N.P. Ong, R.J. Cava, Nature 411 (2001) 54.
[2] I.R. Shein, A.L. Ivanovskii, E.Z. Kurmaev, A. Moewes, S. Chiuzbian, L.D. Finkelstein, M. Neumann, Z.A. Ren, G.C. Che, Phys. Rev. B 66 (2002) 024520.
[3] A.L. Ivanovskii, Phys. Solid State 45 (2003) 1829.
[4] S. Mollah, J. Phys.: Cond. Matter 16 (2004) R1237.
[5] M. Uehara, T. Yamazaki, T. Kori, T. Kashida, Y. Kimishima, I. Hase, J. Phys. Soc. Jpn. 76 (2007) 034714.
[6] V.V. Bannikov, I.R. Shein, A.L. Ivanovskii, Phys. Solid State 49 (2007) 1704.
[7] I.R. Shein, V.V. Bannikov, A.L. Ivanovskii, Physica C 468 (2008) 1.
[8] M. Uehara, A. Uehara, K. Kozawa, T. Yamazaki, Y. Kimishima, Physica C 470 (2010) S688.
[9] V.V. Bannikov, I.R. Shein, A.L. Ivanovskii, Physica B 405 (2010) 4615.
[10] C.M.I. Okoye, Physica B 405 (2010) 1562.
[11] V.V. Bannikov, I.R. Shein, A.L. Ivanovskii, J. Struct. Chem. 51 (2010) 170.
[12] M.A. Helal, A.K.M.A Islam, Physica B 406 (2011) 4564.
[13] J. Winterlik, G.H. Fecher, C. Felser, M. Jourdan, K. Grube, F. Hardy, H. von Löhneysen, K.L. Holman, R. J. Cava, Phys. Rev. B 78 (2008) 184506.
[14] F. Ronning, E. D. Bauer, T. Park, N. Kurita, T. Klimczuk, R. Movshovich, A.S. Sefat, D. Mandrus, J.D. Thompson, Physica C 469 (2009) 396.
[15] Y. Matsumura, H. Ogino, S. Horii, Y. Katsura, K. Kishio, J.-I. Shimoyama, Appl. Phys. Express 2 (2009) 063007.
[16] A.L. Ivanovskii, Russ. Chem. Rev. 79 (2010) 1.
[17] I.R. Shein, D.V. Suetin, A.L. Ivanovskii, Physica B 406 (2011) 676.
[18] A.J.S. Machado, T. Grant, Z. Fisk, Supercond. Sci. Technol. 24 (2011) 095007.
[19] A.J.S. Machado, T. Grant, Z. Fisk, Supercond. Sci. Technol. 25 (2012) 045010.
[20] M.A. Moss, W. Jeitschko, Z. Anorgan Allgem. Chem. 603 (1991) 57.
[21] P. Blaha, K. Schwarz, G. Madsen, D. Kvasnicka D, J. Luitz, WIEN2k, An Augmented Plane Wave Plus Local Orbitals Program for Calculating Crystal Properties, Vienna University of Technology, Vienna (2001).
[22] J.P. Perdew, S. Burke, M. Ernzerhof, Phys. Rev. Lett. 77 (1996) 3865.
[23] A.H. MacDonald, W.E. Pickett, D.D. Koelling, J. Phys. C 13 (1980) 2675.
[24] I.R. Shein, A.L. Ivanovskii, Phys. Solid State 52 (2010) 2039.
[25] I.S. Lim, G.E. Scuseria, Chem. Phys. Lett., 460 (2008) 137.
[26] H. Kleykamp, Thorium Carbides. Gmelin Handbook of Inorganic and Organometallic Chemistry, eighth ed. Thorium supplement, Vol. C6, Springer, Berlin (1992).
[27] I.R. Shein, K.I. Shein, A.L. Ivanovskii, J. Nuclear Mater. 353 (2006) 19.
[28] I.R. Shein, A.L. Ivanovskii, J. Struct. Chem. 49 (2008) 348.
[29] A.L. Ivanovskii, Progr. Mater. Sci. 51 (2012) 184.
[30] M. Born, K. Huang, Dynamical Theory of Crystal Lattices. Oxford, Clarendon (1956).
[31] S.F. Pugh, Philos. Mag. 45 (1953) 823.
[32] S.Y. Chen, X.G. Gong, S.H. Wei, Phys. Rev. B 77 (2008) 014113.
[33] S.I. Ranganathan, M. Ostoja-Starzewski, Phys. Rev. Lett. 101 (2008) 055504.
[34] H. Chung, W.R. Buessem, In: Anisotropy in Single Crystal Refractory Compound, Eds. F.W. Vahldiek, S.A. Mersol, Plenum Press (1968)
[35] I.R. Shein, K.I. Shein, N.I. Medvedeva, A.L. Ivanovskii, phys. status solidi (b), 244 (2007) 3198.
[36] R.F.W. Bader, Atoms in Molecules: A Quantum Theory. International Series of Monographs on Chemistry, Clarendon Press, Oxford, (1990)